# Suppression of Superconductivity in Heavy-ion Irradiated 2*H*-NbSe$_2$ Caused by Negative Pressure

Wenjie Li[1*], Sunseng Pyon[1], Ataru Ichinose[2], Satoru Okayasu[3], and Tsuyoshi Tamegai[1]

[1]*Department of Applied Physics, The University of Tokyo, 7-3-1 Hongo, Bunkyo-ku, Tokyo 113-8656, Japan*

[2]*Grid Innovation Research Laboratory, Central Research Institute of Electric Power Industry, 2-6-1 Nagasaka, Yokosuka, Kanagawa 240-0196, Japan*

[3]*Advanced Science Research Center, Japan Atomic Energy Agency, Tokai, Ibaraki 319-1195, Japan*

Effects of columnar defects created by 320 MeV Au irradiation on 2*H*-NbSe$_2$ single crystals with a dose equivalent matching field ($B_\Phi$) up to 16 T were studied. Critical temperature ($T_c$) is found to be suppressed almost linearly at a rate of 0.07 K/T. At the same time, the lattice parameters *a* and *c* are found to be expanded at rates of 0.016%/T and 0.030%/T, respectively. Such a lattice expansion should work as *negative pressure* to affect $T_c$. By separating the effect of heavy-ion irradiation on $T_c$ suppression through lattice expansion and disorder, it is found that $T_c$ is suppressed more by lattice expansion rather than by disorder.

## 1. Introduction

Particle irradiation is one of the most effective methods to enhance the critical current density ($J_c$) of superconductors. The enhancement of $J_c$ has been demonstrated by many experiments in cuprate [1-7] and iron-based superconductors [8-14], and corresponding mechanisms have been proposed theoretically [15]. In addition, it has also been reported that the artificial defects introduced by heavy-ion irradiations can act as disorder and suppress $T_c$ [16-18]. However, most of the particle irradiation experiments have been conducted on unconventional high-temperature superconductors, with a few exceptions done on conventional superconductors. Among all the conventional superconductors, NbSe$_2$ is one of such superconductors, on which effects of heavy-ion irradiations have been studied. For example, the condition for the creation of columnar defects in NbSe$_2$ has been discussed by irradiating samples with different energy of heavy-ions [19]. Dynamics of vortices [20,21], lock-in phenomenon [22], and vortex creep [23] in NbSe$_2$ with columnar defects introduced by heavy-ion irradiations have been discussed. However, as far as we know, there have been no systematic study on how $T_c$ is affected by heavy-ion irradiations in NbSe$_2$.

As is known from the Anderson theorem, nonmagnetic disorder will not affect $T_c$ if a superconductor has an *isotropic s*-wave gap structure [24]. Although NbSe$_2$ is known to be an *s*-wave superconductor, thermal transport [25], angle-resolved photoemission spectroscopy [26], and scanning tunneling microscope measurements [27] have revealed that it has an *anisotropic s*-wave gap structure, which means that nonmagnetic disorder can affect the $T_c$ in NbSe$_2$. For example, the disorder induced $T_c$ suppression has been observed in a previous study about the effects of 2.5 MeV electron irradiation on NbSe$_2$ [28]. However, a slight initial enhancement of $T_c$ with increasing irradiation dose has also been observed before the $T_c$ started to decline. The initial enhancement of $T_c$ is attributed to the progressive suppression of charge-density wave (CDW) that competes with superconductivity. While CDW transition in NbSe$_2$ is weak compared with other prominent CDW transitions such as observed in NbSe$_3$ [29], the interplay between superconductivity and CDW makes interpretation of the irradiation effect in NbSe$_2$ complex.

On the other hand, application of pressure is an effective way to explore the underlying properties of a material (such as CDW and superconductivity). Actually, there have been many studies reporting the enhancement of $T_c$ in superconductors after the application of hydrostatic pressure [30-34]. At the same time, hydrostatic pressure always works to shrink the lattice parameters [34-37]. These results indicate that the $T_c$ enhancement can be attributed to the shrinkage of the lattice. In general, hydrostatic pressure can only provide *positive pressure* on superconductors. However, if we can expand lattice, it is considered to be equivalent to *negative pressure*. Since it has been reported that particle irradiation can expand lattice in various materials [38-42], it is necessary to inspect the effect of heavy-ion irradiation to get some insight into the interplay between $T_c$ suppression and the lattice expansion in NbSe$_2$.

In the present paper, we report systematic studies on the suppression of $T_c$ in NbSe$_2$ upon heavy-ion irradiation and discuss how $T_c$ is affected by the lattice expansion, disorder, and CDW.

## 2. Experimental Details

Single crystals of 2*H*-NbSe$_2$ were prepared by the iodine vapor transport method [43]. Stoichiometric amounts of Nb powder (3N) and Se shot (5N), 2 g in total, were sealed in an evacuated quartz tube (16 mm inner diameter and 200 mm length) with 5 mg/cm$^3$ of iodine.

The evacuated quartz tube was kept in a tube furnace with a temperature gradient of ~50 ℃ (high temperature at 800 ℃ and low temperature at 750 ℃). The vapor transport process lasted for two weeks, followed by furnace cooling.

The 320 MeV Au irradiation experiments were performed at JAEA in Tokai. The 800 MeV Xe irradiations were performed at NIRS-HIMAC in Chiba. Before starting the irradiation experiment, the single crystals were prepared to thin plates with a thickness less than 10 $\mu$m along the $c$-axis, which is thinner than the projected range of 320 MeV Au and 800 MeV Xe irradiations for $NbSe_2$ of ~17 $\mu$m and ~35 $\mu$m, respectively. The projected ranges are calculated by SRIM-2008 (the Stopping and Range of Ions in Matter-2008) [44]. For creating columnar defects parallel to the $c$-axis up to a dose equivalent matching field of $B_\Phi = 16$ T, the incident ions were irradiated along the $c$-axis. In the case of 800 MeV Xe, some crystals were irradiated in a splayed fashion with splay angles of ±20 ° from the $c$-axis to confirm the presence of columnar defects. The lattice parameters before and after the irradiation were characterized by X-ray diffraction (XRD) measurements using a commercial diffractometer (Smartlab, Rigaku) with Cu $K\alpha$ radiation. The Bragg peaks other than (00$l$) peaks cannot be directly observed by using standard $\omega$-2$\theta$ scan ($\omega$: sample angle, 2$\theta$: detector angle) XRD measurements. According to the Laue equation, the condition for the constructive interference is

$$\Delta \boldsymbol{k} = \boldsymbol{k}_{\text{out}} - \boldsymbol{k}_{\text{in}} = \boldsymbol{G}, \qquad (1)$$

where $\boldsymbol{k}_{\text{out}}$, $\boldsymbol{k}_{\text{in}}$, and $\boldsymbol{G}$ are outgoing and incoming wave number of X-ray, and reciprocal lattice vector, respectively. To make ($h$0$l$) peaks observable, ($h$0$l$) reciprocal lattice point has to be brought onto the Ewald sphere by rotating the crystal. Through successive optimizations of $\omega$ and 2$\theta$, we can determine 2$\theta$ for ($h$0$l$) peaks, from which we can determine $d$-spacing for ($h$0$l$) plane. Before the calculation, we first fitted all Bragg peaks by using the extended Gaussian functions to account for the presence of two wave lengths of X-ray. The $c$ lattice parameter is calculated by taking an average value from (004), (006), and (008) peaks for better accuracy. For the calculation of $a$ lattice parameter, we observed six equivalent ($10\bar{1}0$) peaks and took their average. After determining diffraction angles, Bragg's law was used to calculate corresponding $d$-spacings, which is related to $a$ lattice parameter by the relation for the hexagonal lattice of

$$\frac{1}{d^2} = \frac{4(h^2 + hk + k^2)}{3a^2} + \frac{l^2}{c^2}, \qquad (2)$$

where $h$, $k$, and $l$ are Miller indices. The in-plane resistivity was measured by AC four-probe method using AC resistance bridge (LR700, Linear Research). For the magneto-optical (MO) measurement, an iron-garnet indicator film was placed in contact with the thin $NbSe_2$ single crystal and cooled down with a He-flow cryostat (Microstat-HR, Oxford Instruments). The MO image was taken by using a cooled-CCD camera with 12-bit resolution (ORCA-ER, Hamamatsu). Cross-sectional observations of the irradiated $NbSe_2$ were performed with a high-resolution scanning transmission electron microscope (TEM, JEOL, JEM-3000F). Magnetization measurements were performed using a commercial SQUID magnetometer (MPMS-XL5, Quantum Design). The extended Bean critical state model [45] is used to estimate the $J_c$ after measuring the magnetic hysteresis loops (MHLs). In this model $J_c$ (A/cm$^2$) is given by

$$J_c = \frac{20\Delta M}{a\left(1-\dfrac{a}{3b}\right)} \quad (a<b), \tag{3}$$

where $\Delta M$ (emu/cm$^3$) is the difference of magnetization when sweeping the external field down and up. $a$ (cm) and $b$ (cm) are the width and length of the sample, respectively.

## 3. Results and Discussion

### 3.1 Creation of columnar defects

Figure 1 (a) shows a TEM image of the 320 MeV Au irradiated $NbSe_2$ single crystal with $B_\Phi$ = 8 T, where columnar defects with diameter of 6~8 nm are observed clearly. Figure 1 (b) shows a TEM image of a $NbSe_2$ single crystal irradiated by 800 MeV Xe from two directions with $\theta_{CD}$ = ±20 ° from the $c$-axis, each with $B_\Phi$ = 2 T, where columnar defects are not so clear, but some faint traces of columnar defects with diameter of 4~6 nm are still observed. The reason why we put the TEM results on $NbSe_2$ irradiated by 800 MeV Xe from two directions rather than those parallel to $c$-axis is to confirm that dark contrasts we observed here are really irradiation induced defects. The difference between the defect structures created by two kinds of ions with different energy can be explained by the difference in the stopping power of these ions. The electronic stopping power $S_e$ has been used as one of reference conditions to predict whether the columnar defects can be created or not. Empirically, $S_e$ = 2 keV/Å is accepted as the threshold for the formation of columnar defects in cuprate superconductors [46]. $S_e$ value for 320 MeV Au irradiation and 800 MeV Xe irradiation into $NbSe_2$ are calculated as $S_e$ = 3.27 keV/Å and $S_e$ = 2.74 keV/Å using SRIM-2008, and columnar defects were observed in both of these two cases. Previous studies on heavy-ion irradiations of 1 GeV Pb ($S_e$ = 3.98 keV/Å) and 500 MeV I ($S_e$ = 2.78 keV/Å) on $NbSe_2$ single crystals also clarified the presence of columnar defects [19], while in the case of 230 MeV Ni ($S_e$ = 1.17 keV/Å) irradiation on NbSe$_2$, no columnar defects were found [19]. One point we need to note here is that no clear columnar defects were observed in NbSe$_2$ irradiated by 340 MeV Xe ($S_e$ = 2.69 keV/Å > 2 keV/Å) [19]. From these observations, we can conclude that the threshold for the creation of columnar defects in NbSe$_2$ is close to $S_e$ = 2.7 keV/Å. Figure 1 (c) shows an MO image of NbSe$_2$ after the irradiation in the remanent state at $T$ = 5 K. It shows the presence of a clear double-Y shaped current discontinuity line and confirms that uniform persistent current is flowing following the rectangular shape of the crystal. Actually, when a homogeneous rectangular superconductor is driven into the critical state, a uniform current parallel to the edge of the sample flows, forming a characteristic double Y-shaped current discontinuity lines as shown in Fig. 1 (d).

*3.2 Effects of columnar defects on lattice parameters*

Figures 2 (a) and (b) show (006) and (10$\bar{1}$0) Bragg peaks for NbSe$_2$ irradiated by 320 MeV Au with different $B_\Phi$, where the Bragg peaks shift monotonically from higher angles to lower angles with increasing $B_\Phi$. As a reference, we also show (006) and (10$\bar{1}$0) Bragg peaks for NbSe$_2$ irradiated by 800 MeV Xe in Figs. 2 (c) and (d), respectively, where similar shifts of Bragg peaks are observed although their magnitudes are smaller compared with the case of 320 MeV Au irradiation. Lattice parameters *a* and *c*, calculated by the method as described above, as functions of $B_\Phi$ are shown in Figs. 3 (a) and (b), respectively. For NbSe$_2$ irradiated by 320 MeV Au, *a* and *c* increase linearly up to $B_\Phi$ = 16 T at rates of 0.016%/T and 0.030%/T, respectively. Similarly, for NbSe$_2$ irradiated by 800 MeV Xe, *a* and *c* increase linearly up to $B_\Phi$ = 16 T at smaller rates of 0.0037%/T and 0.0038%/T, respectively. The lattice expansion for 800 MeV Xe irradiated samples is much smaller than that for 320 MeV Au irradiated samples. The large difference is most probably originated from the different morphology of the defects. As can be seen in Figs. 1 (a) and (b), clear columnar defects have been successfully introduced in NbSe$_2$ by 320 MeV Au irradiation, while only faint traces of columnar defects are observed in samples irradiated by 800 MeV Xe. Figs. 3 (c) and (d) are the FWHM (full width at half maximum) values of XRD peaks for NbSe$_2$ single crystals before and after the irradiations by 320 MeV Au and 800 MeV Xe. The FWHM shows almost no change after heavy-ion irradiations, which indicates that the lattice expands uniformly by irradiation. The reason why the lattice parameters expanded is because the amorphous materials created by rapid cooling of melted crystal have lower density with larger separation of constituent atoms. Due to expansion of the volume of columnar defects, surrounding lattice parts expand both along columnar defects and perpendicular to them. The expansion of the lattice parameters due to particle irradiations has also been widely observed in various materials such as $La_{1.85}Sr_{0.15}CuO_4$ [38], $CeO_2$ [39], $MgB_2$ [40], $YBa_2Cu_3O_{7-y}$ [41], and $ErBa_2Cu_3O_7$ [42]. For example, the *c*-axis lattice parameter expansion in $ErBa_2Cu_3O_7$ thin film has been reported at rates of 0.021%/T (3.8 GeV Ta irradiation), 0.050%/T (80 MeV I irradiation), 0.081%/T (200 MeV I irradiation), and 0.16%/T (120 MeV Au irradiation) [42]. These values compare well with 0.030%/T for 320 MeV Au irradiated $NbSe_2$ crystals.

*3.3 Effects of columnar defects on $T_c$*

The inset of Fig. 4 (a) shows *M - T* curves for $NbSe_2$ irradiated by 320 MeV Au with different $B_\Phi$. It is clear that as $B_\Phi$ increases, $T_c$ decreases monotonically. Figure 4 (a) shows the $B_\Phi$ dependence of $T_c$ in these samples, where $T_c$ is determined by the onset of diamagnetism in *M-T* curves. As a reference, similar results obtained for 800 MeV Xe irradiated $NbSe_2$ are also plotted [47]. $T_c$ decreases significantly in $NbSe_2$ after irradiation of 320 MeV Au. Such appreciable suppression of $T_c$ upon introduction of columnar defects by 320 MeV Au looks strange since the defects are concentrated only along linear tracks leaving other area intact. One important fact that should be noted here is that during the process of creation of columnar defects, energetic secondary electrons are generated and those electrons can create atomic scale point defects, which are not easy to be observed by using TEM. However, there is another important factor that may affect $T_c$ upon heavy-ion irradiations, which is the lattice expansion induced by the creation of columnar defects. As we mentioned in the last section, both *a*-axis and *c*-axis lattice parameters for $NbSe_2$ after 320 MeV Au irradiation show monotonic increase. Namely, introduction of columnar defects via heavy-ion irradiation is equivalent to the application of *negative pressure*, which is not easy to realize in a real material. Here, we try to compare the $T_c$ variation by the lattice parameter change induced by heavy-ions irradiations and by pressure.

To calculate how much $T_c$ is affected by lattice parameter change induced by irradiation, we separate effects related to change in *a*-axis from that in *c*-axis. Based on the Ehrenfest relation, the uniaxial pressure dependence of $T_c$ can be calculated by

$$\frac{dT_c}{dp_i} = \frac{3V\Delta\alpha_i T_c}{\Delta C_p} \quad (i = a,b,c), \tag{4}$$

where *V* is the specific volume, $\Delta\alpha_i$ and $\Delta C_p$ are changes in thermal expansion coefficient and heat capacity across $T_c$ [48]. Using experimental data of $V$, $\Delta \alpha_i$, and $\Delta C_p$ for NbSe$_2$, $dT_c/dp_c$ = -0.145 K/kbar is calculated in Ref. [48], which is close to the experimental result of -0.14 K/kbar where the uniaxial pressure was applied parallel to the $c$-axis [49]. The hydrostatic pressure derivative of $T_c$ is related to the uniaxial pressure derivative of $T_c$, and it can be written as

$$\frac{dT_c}{dp} = \frac{2dT_c}{dp_a} + \frac{dT_c}{dp_c}. \tag{5}$$

By using reported data of $dT_c/dp_c$ [49] and $dT_c/dp$ for NbSe$_2$ [35], $dT_c/dp_a$ is evaluated as 0.100 K/kbar. In Ref. [35], linear isothermal compressibilities in the basal plane and along the $c$-axis are calculated as $K_a = \Delta a/a/\Delta p$ = 4.1 ($\pm$0.4) $\times 10^{-4}$ /kbar and $K_c = \Delta c/c/\Delta p$ = 16.2 ($\pm$0.5) $\times 10^{-4}$ /kbar, respectively. By combining these $K_a$ and $K_c$ values with the lattice parameter data for NbSe$_2$ irradiated by 320 MeV Au at $B_\Phi$ = 16 T, $T_c$ is expected to change -0.62 K (based on $K_a$) and 0.42 K (based on $K_c$), resulting in total $T_c$ change of -0.82 K. This expected $T_c$ suppression is smaller than the experimental value of -1.13 K. If we simply separate the effect of irradiation on $T_c$ into that via lattice expansion and via disorder for NbSe$_2$ irradiated by 320 MeV Au at $B_\Phi$ = 16 T, $T_c$ variation induced by the disorder is -0.31 K. This suggests that $T_c$ in NbSe$_2$ after the introduction of columnar defects is mainly suppressed by lattice expansion rather than by disorder. Similar analyses for 800 MeV Xe irradiated NbSe$_2$ ($B_\Phi$ = 16 T) give $\Delta T_c$ ($a$-axis) = -0.14 K and $\Delta T_c$ ($c$-axis) = 0.05 K, with total $\Delta T_c$ of -0.23 K, which is close to the measured $T_c$ change of -0.23 K. To confirm that the irradiation is properly done in samples irradiated by 800 MeV Xe, we show the $J_c$ - $H$ curves of the irradiated samples in Fig. 4 (b). It is clear that both 320 MeV Au and 800 MeV Xe irradiations enhance $J_c$ significantly compared with the pristine sample. Then, it is also suggested that the effect of 800 MeV Xe irradiation on $T_c$ in NbSe$_2$ is dominated by the lattice expansion rather than disorder. However, since effects of 800 MeV Xe irradiation on both $T_c$ and lattice expansion are much weaker than those of 320 MeV Au irradiation, studies on samples with much larger $B_\Phi$ are necessary to confirm the effect of *negative pressure* more quantitatively.

It should be note that there is additional complication in the case of 2$H$-NbSe$_2$ to discuss the variation of $T_c$ by any perturbation, which is the presence of CDW. It is commonly observed that in superconductors with competing CDW, suppression of CDW will help to enhance the superconductivity. Because the Fermi surface is used to form both superconductivity and CDW, if the CDW is suppressed and a part of the gapped Fermi surface can be released, this part of Fermi surface will be used to enhance the superconductivity. For example, the competition between CDW and superconductivity has been studied by applying hydrostatic pressure in $Lu_5Ir_4Si_{10}$ [50], $CsV_3Sb_5$ [51] as well as $NbSe_2$ [30]. The competition between CDW and superconductivity has also been observed in terms of particle irradiation in $NbSe_2$ [28] and $Lu_5Ir_4Si_{10}$ [52]. Figure 5 (a) shows the $\rho$-$T$ curves for a pristine $NbSe_2$ single crystal and that after irradiated by 320 MeV Au with $B_\Phi = 2$ T, where $T_c$ is suppressed from 7.15 K to 7.04 K, and at the same time $T_{CDW}$ is suppressed from 33 K to 30 K. For the sample irradiated by 320 MeV Au with $B_\Phi = 2$ T, the global feature of $\rho$-$T$ curve and the value of resistivity ($\rho(8\ K) = 7\ \mu\Omega$ cm) and $T_{CDW}$ (~30 K) are similar to that irradiated by 2.5 MeV electron with a dose of 0.47 C/cm$^2$, $\rho(8\ K) = 7.5\ \mu\Omega$ cm and $T_{CDW}$ (~30 K) [28]. However, in the latter case, $T_c$ was increased to 7.45 K.

To discuss why the initial $T_c$ enhancement is not observed in $NbSe_2$ irradiated by 320 MeV Au, one factor that needs to be considered is the sample quality. It has been reported that the CDW transition is sensitively affected by the sample quality [53]. If the sample itself has enough impurities that already suppressed the $T_c$, the resistivity value of the pristine sample should be large, which is not the case as we pointed out above. Furthermore, residual resistivity ratio (RRR) value of $\rho(300\ K)/\rho(8\ K) = 50.7$ of the pristine sample used in the present experiment is very similar to that used in Ref. [28]. These facts indicate that the absence of initial $T_c$ enhancement here is not related to the sample quality. Another possible explanation is that the enhanced $T_c$ by suppression of CDW in the heavy-ion irradiated sample is smaller compared to the total effects from lattice expansion and disorder. As shown schematically in Fig. 5 (b), $T_c$ is monotonically suppressed by the lattice expansion as discussed above. The degree of disorder should also be proportional to $B_\Phi$, which induces a monotonic suppression of $T_c$ with increasing $B_\Phi$. For the effects of CDW, it initially enhances $T_c$. However, after CDW is completely suppressed at $B_\Phi^*$, the effect of CDW on $T_c$ does not change. By the combination of all these three factors, $T_c$ will continuously decrease with increasing $B_\Phi$. Whether we have reached $B_\Phi^*$ or not in the present study needs confirmation by resistivity and/or low-temperature X-ray measurements. If the $T_c$ suppression via lattice expansion is negligible in the case of high-energy electron irradiation, a weak initial enhancement of $T_c$ can be realized as observed in Ref. [28]. Weaker effects of high-energy electron irradiation to the lattice is quite probable considering the fact that created defects are mainly Frenkel pairs which are supposed to affect the lattice only weakly.

## 4. Conclusions

In summary, we have successfully introduced columnar defects in $NbSe_2$ single crystals by

320 MeV Au irradiation. The lattice parameters *a* and *c* are found to be expanded with increasing $B_\Phi$. This lattice expansion can be considered to be equivalent to the *negative pressure* to the lattice. When the effects of 320 MeV Au irradiation on $T_c$ are separated into that via lattice expansion and that via disorder, $T_c$ is found to be affected more by lattice expansion rather than disorder. However, the presence of CDW in NbSe$_2$ makes this interpretation of the effect of irradiation on $T_c$ complicated. To avoid this complication, we are planning to do similar measurements in 2*H*-NbS$_2$ that shows superconductivity around 6.0 K but does not form CDW at any temperatures.

**Acknowledgment**

This work was partially supported by a Grant in Aid for Scientific Research (A) (17H01141) from the Japan Society for the Promotion of Science (JSPS).

*E-mail: wenjiecd@gmail.com

1) L. Civale, A. D. Marwick, T. K. Worthington, M. A. Kirk, J. R. Thompson, L. Krusin-Elbaum, Y. Sun, J. R. Clem, and F. Holtzberg, Phys. Rev. Lett. **67**, 648 (1991).
2) R. C. Budhani, M. Suenaga, and S. H. Liu, Phys. Rev. Lett. **69**, 3816 (1992).
3) Th. Schuster, M. V. Indenbom, H. Kuhn, H. Kronmuller, M. Leghissa, and G. Kreiselmeyer, Phys. Rev. B **50**, 9499 (1994).
4) C. J. van der Beck, M. Konczykowski, V. M. Vinokur, T. W. Li, P. H. Kes, and G. W. Crabtree, Phys. Rev. Lett. **74**, 1214 (1995).
5) M. Sato, T. Shibauchi, S. Ooi, T. Tamegai, and M. Konczykowski, Phys. Rev. Lett. **79**, 3759 (1997).
6) E. Mezzetti, R. Gerbaldo, G. Ghigo, L. Gozzelino, B. Minetti, C. Camerlingo, A. Monaco, G. Cuttone, and A. Rovelli, Phys. Rev. B **60**, 7623 (1999).
7) G. Fuchs, K. Nenkov, G. Krabbes, R. Weinstein, A. Gandini, R. Sawh, B. Mayes, and D. Parks, Supercond. Sci. Technol. **20**, S197 (2007).
8) Y. Nakajima, Y. Tsuchiya, T. Taen, T. Tamegai, S. Okayasu, and M. Sasase, Phys. Rev. B **80**, 012510 (2009).
9) R. Prozorov, M. A. Tanatar, B. Roy, N. Ni, S. L. Budko, P. C. Canfield, J. Hua, U. Welp, and W. K. Kwok, Phys. Rev. B **81**, 094509 (2010).
10) L. Fang, Y. Jia, C. Chaparro, G. Sheet, H. Claus, M. A. Kirk, A. E. Koshelev, U. Welp, G. W. Crabtree, W. K. Kwok, S. Zhu, H. F. Hu, J. M. Zuo, H.H. Wen, and B. Shen, Appl. Phys. Lett. **101**, 012601 (2012).
11) L. Fang, Y. Jia, V. Mishra, C. Chaparro, V. K. VlaskoVlasov, A. E. Koshelev, U. Welp, G. W. Crabtree, S. Zhu, N. D. Zhigadlo, S. Katrych, J. Karpinski, and W. K. Kwok, Nat.

Commun. **4**, 2655 (2013).
12) K. J. Kihlstrom, L. Fang, Y. Jia, B. Shen, A. E. Koshelev, U. Welp, G. W. Crabtree, W. K. Kwok, A. Kayani, S. F. Zhu, and H. H. Wen, Appl. Phys. Lett. **103**, 202601 (2013).
13) N. Haberkorn, J. Kim, K. Gofryk, F. Ronning, A. S. Sefat, L. Fang, U. Welp, W. K. Kwok, and L. Civale, Supercond. Sci. Technol. **28**, 055011 (2015).
14) T. Tamegai, T. Taen, H. Yagyuda, Y. Tsuchiya, S. Mohan, T. Taniguchi, Y. Nakajima, S. Okayasu, M. Sasase, H. Kitamura, T. Murakami, T. Kambara, and Y. Kanai, Supercond. Sci. Technol. **25**, 084008 (2012).
15) D. R. Nelson and V. M. Vinokur, Phys. Rev. B **48**, 13060 (1993).
16) V. Hardy, J. Provost, D. Groult, M. Hervieu, B. Raveau, S. Durcok, E. Pollert, J. C. Frison, J. P. Chaminade, and M. Pouchard, Physica C **191**, 85 (1992).
17) Y. Sun, S. Kittaka, S. Nakamura, T. Sakakibara, K. Irie, T. Nomoto, K. Machida, J. Chen, and T. Tamegai, Phys. Rev. B **96**, 220505 (2017).
18) G. Ghigo, G. A. Ummarino, L. Gozzelino, R. Gerbaldo, F. Laviano, D. Torsello, and T. Tamegai, Sci. Rep. **7**, 13029 (2017).
19) B. Bauer, C. Giethmann, M. Kraus, T. Marek, J. Burger, G. Kreiselmeyer, G. Saemann-lsehenko, and M. Skibowski, Europhys. Lett. **23**, 585 (1993).
20) J. Zhang, L. E. De Long, V. Majidi, and R. C. Budhani, Phys. Rev. B **53**, R8851 (1996).
21) M.-O. André, M. Polichetti, H. Pastoriza, and P. H. Kes, Physica C **338**, 179 (2000).
22) A. V. Silhanek, L. Civale, and M. A. Avila, Phys. Rev. B **65**, 174525 (2002).
23) S. Eley, K. Khilstrom, R. Fotovat, Z. L. Xiao, A. Chen, D. Chen, M. Leroux, U. Welp, W. K. Kwok, and L. Civale, Sci. Rep. **8**, 13162 (2018).
24) P. W. Anderson, J. Phys. Chem. Solids **11**, 26 (1959).
25) E. Boaknin, M. A. Tanatar, J. Paglione, D. Hawthorn, F. Ronning, R. W. Hill, M. Sutherland, L. Taillefer, J. Sonier, S. M. Hayden, and J. W. Brill, Phys. Rev. Lett. **90**, 117003 (2003).
26) T. Yokoya, T. Kiss, A. Chainani, S. Shin, M. Nohara, and H.Takagi, Science **294**, 2518 (2001).
27) H. F. Hess, R. B. Robinson, and J. V. Waszczak, Phys. Rev. Lett. **64**, 2711 (1990).
28) K. Cho, M. Konczykowski, S. Teknowijoyo, M. A. Tanatar, J. Guss, P. Gartin, J. M. Wilde, A. Kreyssig, R. McQueeney, A. I. Goldman, V Mishra, P. J. Hirschfeld, and R. Prozorov, Nat. Commun. **9**, 2796 (2018).
29) A. Briggs, P. Monceau, M. Nunez-Regueiro, J. Peyrard, M. Ribault, and J. Richard, J. Phys. C: Solid State Phys. **13**, 2117 (1980).
30) A. Majumdar, D. VanGennep, J. Brisbois, D. Chareev, A. V. Sadakov, A. S. Usoltsev, M. Mito, A. V. Silhanek, T. Sarkar, A. Hassan, O. Karis, R. Ahuja, and M. Abdel-Hafiez, Phys. Rev. Mater. **4**, 084005 (2020).
31) E. Gati, L. Xiang, S. L. Bud'ko, and P. C. Canfield, Ann. Phys. **532**, 2000248 (2020).
32) A. S. Sefat, Rep. Prog. Phys. **74**, 124502 (2011).

33) F. F. Tafti, A. Juneau-Fecteau, M-È. Delage, S. R. de Cotret, J-Ph. Reid, A. F. Wang, X-G. Luo, X. H. Chen, N. Doiron-Leyraud, and L. Taillefer, Nat. Phys. **9**, 349 (2013).
34) H. Takahashi, A. Sugimoto, Y. Nambu, T. Yamauchi, Y. Hirata, T. Kawakami, M. Avdeev, K. Matsubayashi, F. Du, C. Kawashima, H. Soeda, S. Nakano, Y. Uwatoko, Y. Ueda, T. J. Sato, and K. Ohgushi, Nat. Mater. **14**, 1008 (2015).
35) R. E. Jones, H. R. Shanks, D. K. Finnemore, and B. Morosin, Phys. Rev. B **6**, 835 (1972).
36) W. O. Uhoya, J. M. Montgomery, G. M. Tsoi, Y. K. Vohra, M. A. McGuire, A. S. Sefat, B. C. Sales, and S. T. Weir, J. Phys. Condens. Matter **23**, 122201 (2011).
37) S. Margadonna, Y. Takabayashi, Y. Ohishi, Y. Mizuguchi, Y. Takano, T. Kagayama, T. Nakagawa, M. Takata, and K. Prassides, Phys. Rev. B **80**, 064506 (2009).
38) X. Fan, M. Terasawa, T. Mitamura, T. Kohara, K. Ueda, H. Tsubakino, A. Yamamoto, T. Murakami, and S. Matsumoto, Nucl. Instrum. Methods Phys. Res. B **121**, 331 (1997).
39) Y. Yamamoto, N. Ishikawa, F. Hori, and A. Iwase, Quantum Beam Sci. **4**, 26 (2020).
40) S. G. Jung, S. K. Son, D. Pham, W. C. Lim, J. Song, W. N. Kang, and T. Park, Supercond. Sci. Technol. **34**, 129501 (2021).
41) R. Biswal, J. John, P. Mallick, B. N. Dash, P. K. Kulriya, D. K. Avasthi, D. Kanjilal, D. Behera, T. Mohanty, P. Raychaudhuri, and N. C. Mishra, J. Appl. Phys. **106**, 053912 (2009).
42) A. Iwase, N. Ishikawa, Y. Chimi, K. Tsuru, H. Wakana, O. Michikami, and T. Kambara, Nucl. Instrum. Methods Phys. Res. B **557**, 146 (1998).
43) M. Naito and S. Tanaka, J. Phys. Soc. Jpn. **51**, 219 (1982).
44) J. Ziegler, J. Biersack, and U. Littmark, The Stopping and Range of Ions in Solids (Pergamon, New York, 1985).
45) C. P. Bean, Rev. Mod. Phys. **36**, 31 (1964).
46) Y. Zhu, Z. X. Cai, R. C. Budhani, M. Suenaga, and D. O. Welch, Phys. Rev. B **48**, 6436 (1993).
47) W. Li, S. Pyon, A. Takahashi, D. Miyawaki, Y. Kobayashi, and T. Tamegai, J. Phys.: Conf. Ser. **1590**, 012003 (2020).
48) V. Eremenko, V. Sirenko, V. Ibulaev, J. Bartolomé, A. Arauzo, and G. Reményi, Physica C **469**, 259 (2009).
49) T. Sambongi, J. Low Temp. Phys. **18**, 139 (1975).
50) R. N. Shelton, L. S. Hausermannberg, P. Klavins, H. D. Yang, M. S. Anderson, and C. A. Swenson, Phys. Rev. B **34**, 4590 (1986).
51) F. H. Yu, D. H. Ma, W. Z. Zhuo, S. Q. Liu, X. K. Wen, B. Lei, J. J. Ying, and X. H. Chen, Nat. Commun. **12**, 3645 (2021).
52) M. Leroux, V. Mishra, C. Opagiste, P. Rodière, A. Kayani, W. Kwok, and U. Welp, Phys. Rev. B **102**, 094519 (2020).
53) K. Iwaya, T. Hanaguri, A. Koizumi, K. Takaki, A. Maeda, and K. Kitazawa, Physica B **329**, 1598 (2003).

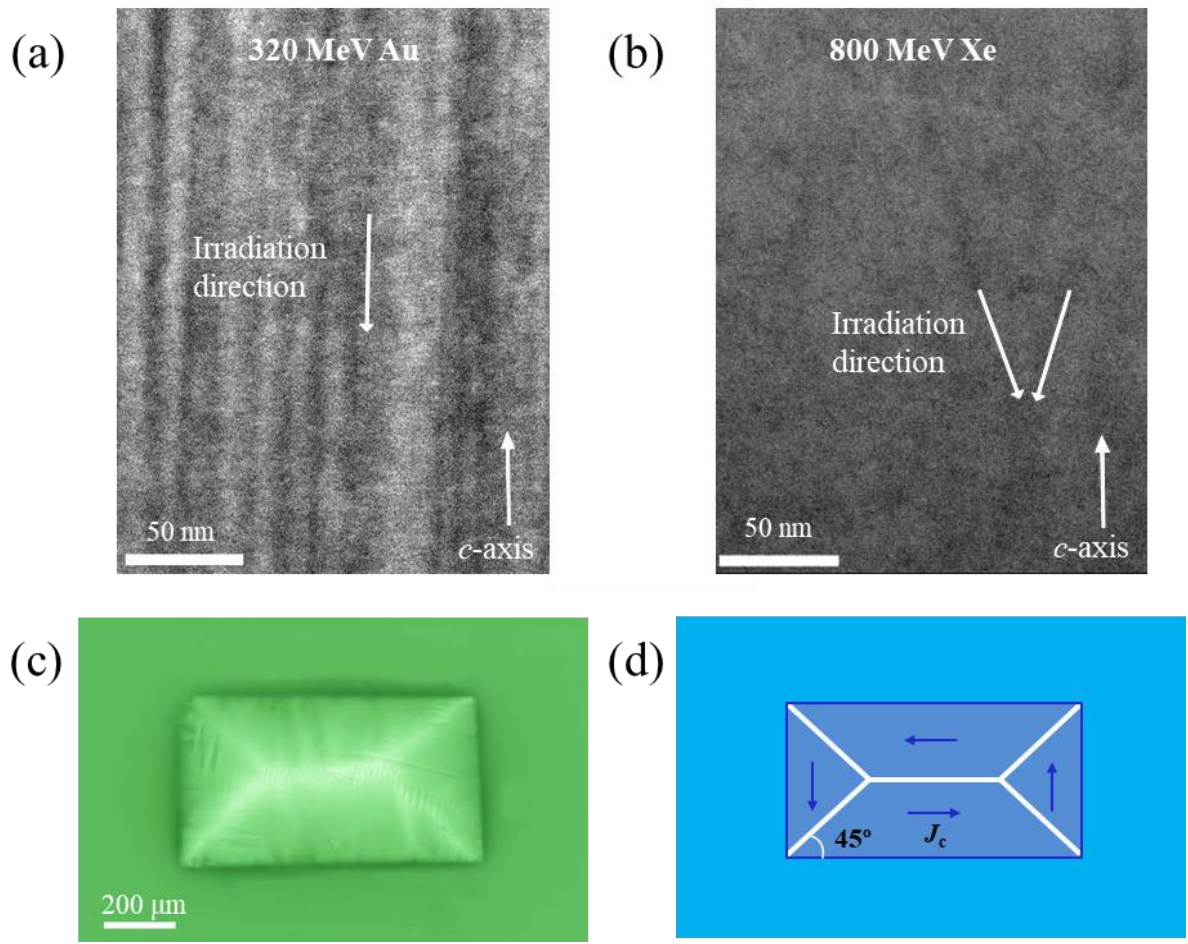

Fig. 1. (a) Transmission electron microscope image for 320 MeV Au irradiated NbSe$_2$ with $B_\Phi$ = 8 T. Clear columnar defects with diameters of 6~8 nm are observed. (b) Transmission electron microscope image for 800 MeV Xe irradiated NbSe$_2$ with $\theta_{CD} = \pm 20\,°$ and $B_\Phi$ = 4 T. Some faint traces of columnar defects with diameter 4~6 nm are observed. (c) Differential MO image of NbSe$_2$ introduced with columnar defects (320 MeV Au, $B_\Phi$ = 2 T) in the remanent state at $T$ = 5 K after cycling the field up to 800 Oe for 0.2 s. (d) Schematic figure of the $J_c$ distribution in a homogeneous rectangular superconductor.

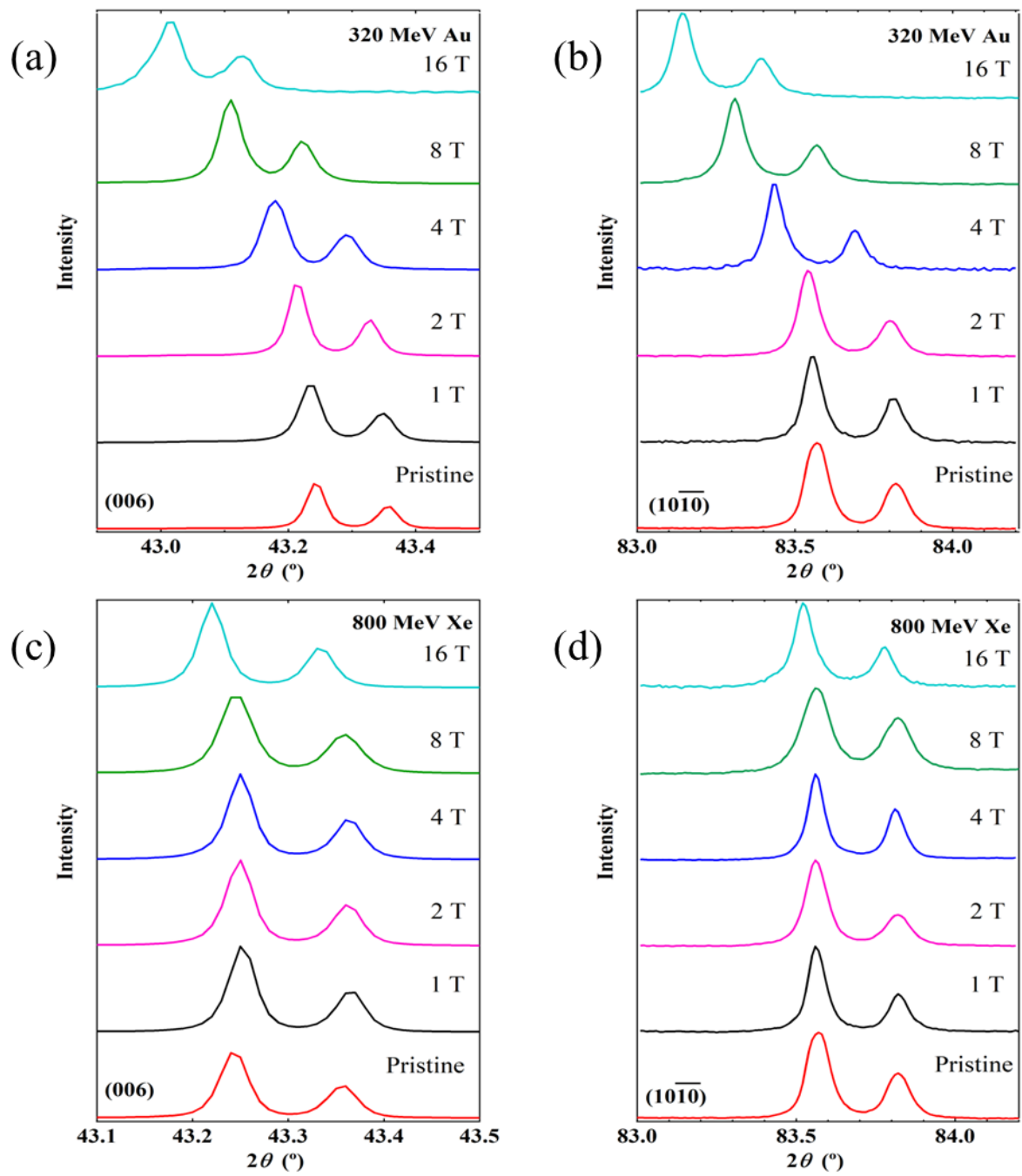

Fig. 2. The XRD patterns of (a) (006) and (b) (10$\bar{1}$0) peaks for NbSe$_2$ single crystals after 320 MeV Au irradiation. Similar data for (c) (006) and (d) (10$\bar{1}$0) peaks taken from ref. [47] are shown for comparison.

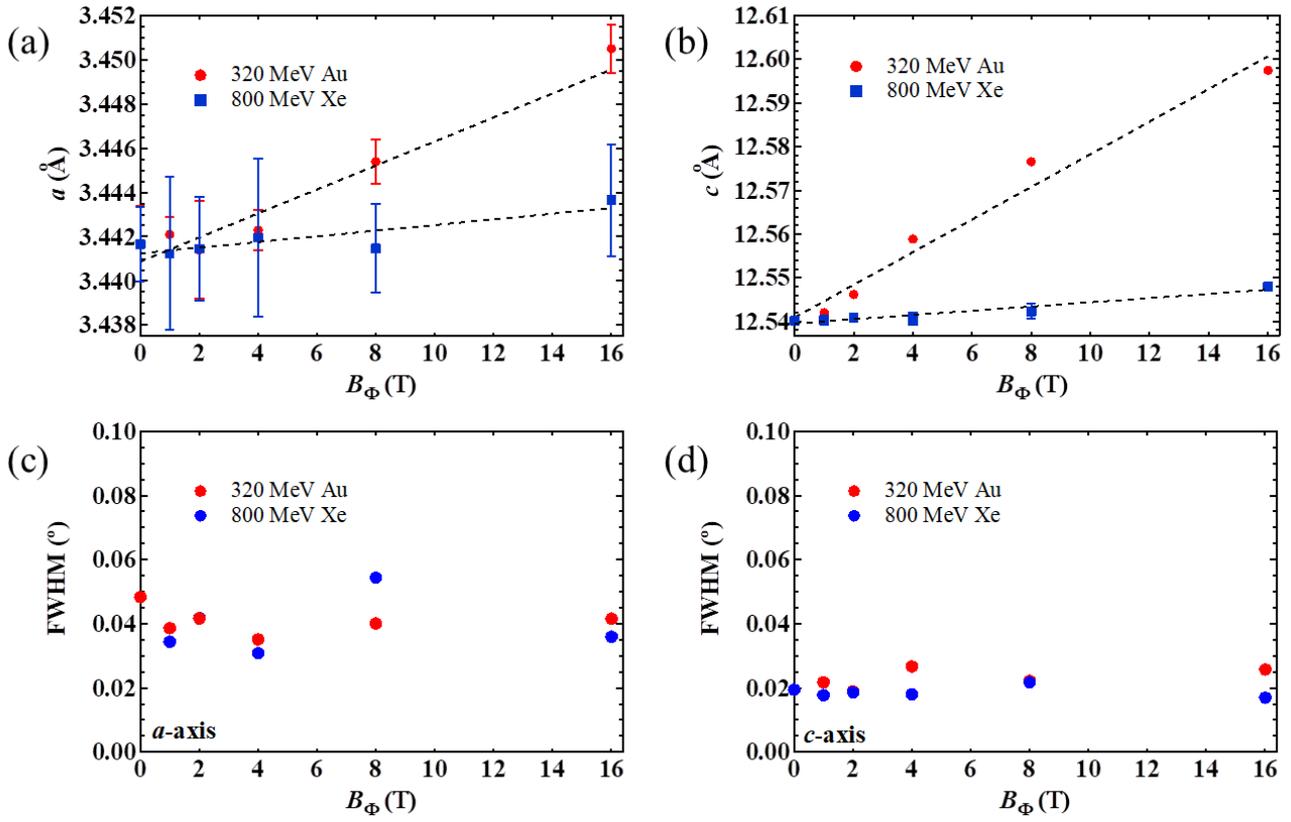

Fig. 3. The $B_\Phi$ dependence of (a) $a$-axis and (b) $c$-axis lattice parameters and the full width at half maximum of XRD peaks for (c) $(10\bar{1}0)$ and (d) $(006)$ peaks for NbSe$_2$ single crystals irradiated by 320 MeV Au and 800 MeV Xe.

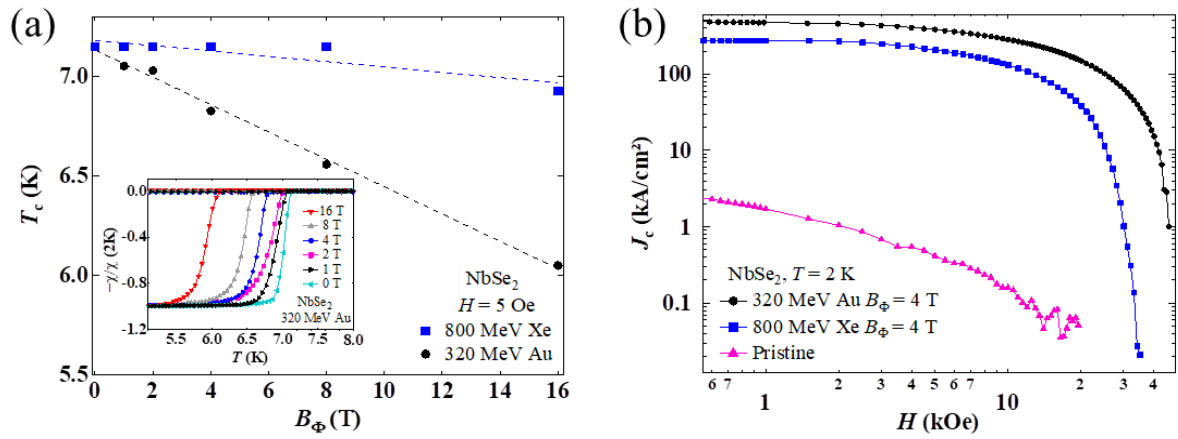

Fig. 4. (a) The $B_\Phi$ dependence of $T_c$ for NbSe$_2$ single crystals after 320 MeV Au irradiation, where a linear suppression of $T_c$ is observed. The data for 800 MeV Xe irradiation are taken from ref. [47]. The inset is the temperature dependence of normalized magnetization at $H = 5$ Oe for NbSe$_2$ single crystals after 320 MeV Au irradiation. (b) The field dependence of $J_c$ for NbSe$_2$ single crystals before and after 320 MeV Au irradiation. The data for 800 MeV Xe irradiation [47] are also included, which confirms that the irradiation is done properly.

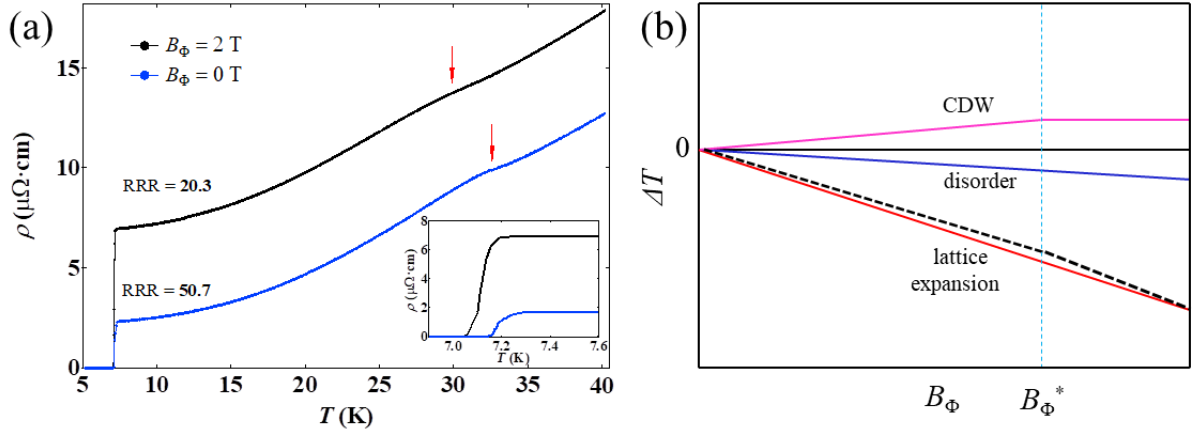

Fig. 5. (a) The temperature dependence of resistivity for NbSe$_2$ single crystals before and after 320 MeV Au irradiation with $B_\Phi$ = 2 T, where $T_{CDW}$ changes from 33 K to 30 K, RRR = $\rho$ ($T$ = 300 K)/ $\rho$ ($T$ = 8 K) changes from 50.7 to 20.3 (CDW transitions are indicated by red arrows). The inset shows a blow-up near the superconducting transition, where $T_c$ is suppressed from 7.15 K to 7.04 K. (b) Schematic changes of $T_c$ variation in NbSe$_2$ induced by lattice expansion, disorder, and CDW. $B_\Phi^*$ is the matching field where CDW is completely suppressed. The combination of these three factors results in the actual change in $T_c$ (black broken line) with increasing $B_\Phi$.